\documentclass[twocolumn]{aastex631}

\usepackage[figuresright]{rotating}
\usepackage{amsmath}
\usepackage{hyperref}
\emergencystretch=\maxdimen
\submitjournal{AJ}

\begin{document}

    \title{LAMOST Medium-resolution Spectroscopic Survey of the Rosette Nebula}

    \correspondingauthor{Chao-Jian Wu, Xuan Fang}
    \email{chjwu@nao.cas.cn, fangx@nao.cas.cn}

    \author{Li-Yue Zhang}
    \affiliation{CAS Key Laboratory of Space Astronomy and Technology, National Astronomical Observatories, Chinese Academy of Sciences, Beijing 100101, China}

    \author{Chao-Jian Wu}
    \affiliation{CAS Key Laboratory of Space Astronomy and Technology, National Astronomical Observatories, Chinese Academy of Sciences, Beijing 100101, China}

    \author{Xuan Fang}
    \affiliation{CAS Key Laboratory of Optical Astronomy, National Astronomical Observatories, Chinese Academy of Sciences, Beijing 100101, China}
    \affiliation{School of Astronomy and Space Sciences, University of Chinese Academy of Sciences, Beijing 100049, China}
    \affiliation{Xinjiang Astronomical Observatory, Chinese Academy of Sciences, 150 Science 1-Street, Urumqi, Xinjiang, 830011, China}
    \affiliation{Laboratory for Space Research, Faculty of Science, The University of Hong Kong, Pokfulam Road, Hong Kong, China}

    \author{Wei Zhang}
    \affiliation{CAS Key Laboratory of Space Astronomy and Technology, National Astronomical Observatories, Chinese Academy of Sciences, Beijing 100101, China}

    \author{Juan-Juan Ren}
    \affiliation{CAS Key Laboratory of Space Astronomy and Technology, National Astronomical Observatories, Chinese Academy of Sciences, Beijing 100101, China}

    \author{Jian-Jun Chen}
    \affiliation{CAS Key Laboratory of Space Astronomy and Technology, National Astronomical Observatories, Chinese Academy of Sciences, Beijing 100101, China}

    \author{Hong Wu}
    \affiliation{CAS Key Laboratory of Space Astronomy and Technology, National Astronomical Observatories, Chinese Academy of Sciences, Beijing 100101, China}

    \begin{abstract}
        We report multi-fiber, medium-resolution spectroscopy of the Rosette Nebula with full spatial coverages, and present a table of the nebular parameters based on the spatially-resolved measurements of emission lines.  These new observations were conducted through the Medium-Resolution Spectroscopic Survey of Nebulae (MRS-N) on the Large Sky Area Multi-Object Fiber Spectroscopic Telescope (LAMOST).  Comprehensive analyses were performed on a total of 3854 high-quality nebular spectra, so far the most extensive spectral dataset available for this nebula that encompasses an area of 4.52 square degrees.  Various physical parameters, including relative line intensities, radial velocities (RVs), and full widths at half maximum (FWHMs), were derived through measurements of the H$\alpha$, [N\,{\sc ii}] $\lambda\lambda$6548,6584 and [S\,{\sc ii}] $\lambda\lambda$6716,6731 emission lines detected in the LAMOST MRS-N spectra.  For the first time, we found a bow-shaped feature in the spatial distribution of RVs of the Rosette Nebula.  Moreover, the spatial distributions of RVs and FWHMs, as well as additional parameters such as gas temperature and turbulent velocity in the vicinity of the nebula, indicate possible interaction between Rosette and the nearby supernova remnant (SNR), Monoceros Loop.  Our new observations provide indispensable measurements of the Rosette Nebula. The parameter table in particular can be used as valuable constraint on the chemo-dynamical modeling of the nebula, which will enable deeper understanding of the characteristics of this H\,{\sc ii} region.
    \end{abstract}


    \section{Introduction} \label{sec:intro}

        H\,{\sc ii} regions are a key composition of the interstellar medium (ISM); they are ionized gas surrounding hot, young and massive stars that excite nebular emission.  The intense radiation emitted by these stars ionizes hydrogen and other molecular constituents, resulting in the expulsion of gas and the formation of a nebula. By analyzing the emission lines from the H\,{\sc ii} regions, researchers can calibrate various physical parameters, including star formation rate, chemical composition, electron temperature, and other relevant parameters. Consequently, the investigation of H\,{\sc ii} regions enhances our understanding of star formation mechanisms and the interactions between high-mass stars and the interstellar medium.

        The Rosette Nebula (NGC\,2237-39, also Caldwell\,49) is situated within the constellation Monoceros, and is one of the most well-known H\,{\sc ii} regions in the Milky Way; it has been extensively studied, both imaging and spectroscopically.  Its distinctive ring-shaped morphology makes it a notable feature in the night sky. The nebula is located approximately 1.55 kpc from Earth \citep{zhao2018} and spans an angular extent of approximately 1.75 degrees. At the core of the nebula lies the central cluster, NGC 2244, which is a young open cluster comprising numerous OB stars that emit substantial ultraviolet radiation and serve as the primary ionization source for the nebula \citep{perez1991}.

        The Monoceros Loop (G205.5+0.5), a prominent supernova remnant (SNR), is situated in the northeastern region of the nebula. This SNR is centered at the right ascension 6$^{h}$ 38$^{m}$ 43$^{s}$ and the declination +06$^{\circ}$ 30$^{\prime}$ 12$^{\prime\prime}$, exhibiting an angular diameter of approximately 220 arcminutes \citep{zhao2018}. The nature of the interaction between the Rosette Nebula and the SNR remains ambiguous. According to Fermi observations, the gamma-ray emission from the SNR is characterized as hard in the area where it intersects with the nebula, suggesting a potential interaction between the two nebulae that may affect the radiation emitted by the gas or the mechanisms involved in star formation. \cite{hardgamma} conducted Fermi-LAT observations of the nebula and discovered that the gamma-ray radiation from the nebula is harder than previously documented, which may be attributable to the influence of the SNR. 

        Because of the extensive size of the nebula, optical spectral observations of the nebula have been infrequent. However, in recent decades, several narrowband observations have been conducted to facilitate a more detailed analysis of the nebula. \cite{meaburn1968} used low-resolution images in H$\alpha$, [O\,{\sc ii}], and [O\,{\sc iii}] to delineate the profile of the dynamical structure. \cite{celnik1983} enhanced the H$\alpha$ imagery with higher resolution, producing an intensity distribution map and calibrating the reddening effects of the nebula by comparing the observation with the observations in the radio band \citep{celnik1986}. Additionally, \cite{RVsample} collected a total of about 700 spectra of the Rosette Nebula, allowing for the calculation of various physical parameters, such as heliocentric velocities, full width at half maximum (FWHM), and electron temperature, although the spatial resolution of these observations was relatively low. Furthermore, \cite{thesis2017} synthesized observational data from multiple projects to conduct a multispectral analysis of the Rosette Nebula; however, the absence of high-quality data for certain significant emission lines, such as the [S\,{\sc ii}] lines, resulted in an incomplete analysis.

        On a more localized level, the substructures in the nebula have been examined comprehensively. In the northwest region of the nebula, dark filaments referred to as ``elephant trunk'' and small dark clumps known as ``globulette'' have been identified, as documented in \cite{globulettes}. Following surface brightness assessments and imaging performed in the infrared band, these structures are believed to be associated with the star formation process, arising from the fragmented shells of molecular clouds \citep{makela2014}. 

        In recent years, the Large Sky Area Multi-Object Fiber Spectroscopic Telescope (LAMOST\footnote{\url{https://www.lamost.org}}, \citealt{{LAMOST1,LAMOST2,LAMOST3,LAMOST4, LAMOST5}}) Medium-Resolution Spectroscopic Survey of Nebulae (MRS-N, \citealt{mrsn}) project has generated a substantial volume of spectrum data of nebulae, thus facilitating a comprehensive understanding of these celestial objects. On the basis of the data from this project, we are now able to perform a thorough spectral analysis for the Rosette Nebula.

        Emission lines from the H\,{\sc ii} regions offer significant insights into the physical processes that occur within the gas. Ultraviolet radiation from protostars and high-mass stars excites atoms, leading to a subsequent recombination process that generates emission lines, such as the H$\alpha$ emission line. As for forbidden lines, including [N\,{\sc ii}] $\lambda\lambda$6548,\,6584, and [S\,{\sc ii}] $\lambda\lambda$6716,\,6731, they are intricately linked to the temperature, electron density, and various other physical conditions present within the nebula \citep{book2006}. Thus, analysis of emission lines within the spectra data is useful in assessing the physical conditions of these regions.

        In this work, Section \ref{sec:methods} introduces the observational data of the Rosette Nebula. Section \ref{sec:results} presents the relative intensity, radial velocity (RV) and full width at half maximum (FWHM) of the nebula as the observational results. Details of the parameter table to be released are also described in this section. Parameters derived from above, including electron density, chemical abundance and ionization fraction, and the special features of the nebula and their possible causes are discussed in Section \ref{sec:discussion}. Finally, conclusions are given in Section \ref{sec:summary}.

    \section{Data} \label{sec:methods}

        \subsection{The LAMOST MRS-N Survey}

            The spectra analyzed in this work were obtained through the LAMOST MRS-N project, which belongs to the LAMOST MRS\footnote{\url{https://dr7.lamost.org/v2.0/doc/mr-data-production-description}} survey carried out earlier \citep{2020arXiv200507210L}.  LAMOST has a circular focal plane with 4000 fibers. In the 5$^\circ$ field of view, each fiber can move within a 2$^{\prime}$ radius. 

            \begin{figure*}[ht!]
            \begin{center}
            \includegraphics[width=17.25cm,angle=0]{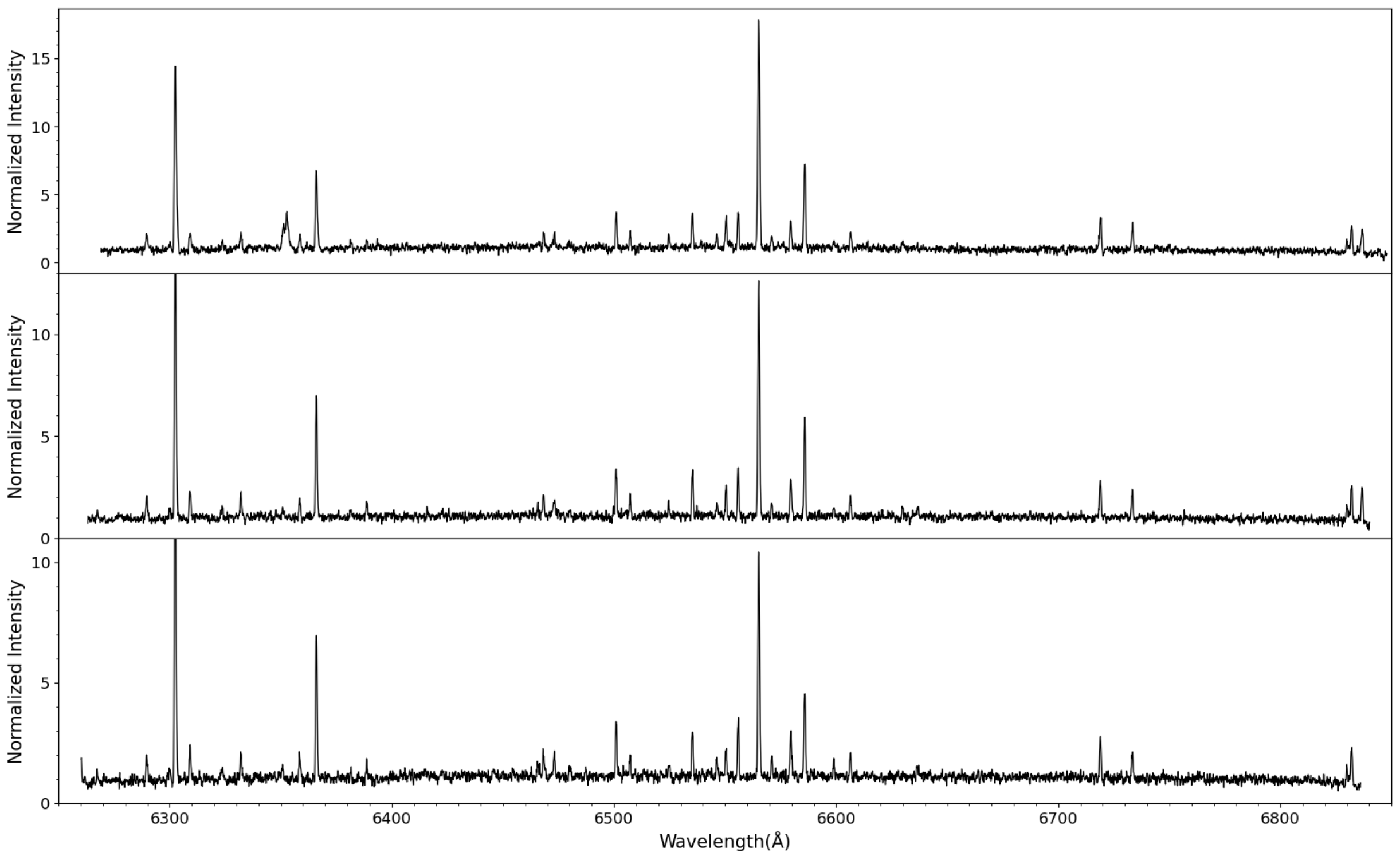}
            \caption{LAMOST medium-resolution spectra of the Rosette Nebula extracted from three different fiber positions, showing the highest data quality with S/N([S\,{\sc ii}])$>$60 (top panel, representing the majority of the fiber spectra reported in this work), medium data quality with S/N([S\,{\sc ii}])$\simeq$40 (middle panel), and relatively low data quality with S/N([S\,{\sc ii}])$\simeq$10.  The spectra are all normalized such that the integrated flux $F$(H$\alpha$) = 300 (given that the theoretical intensity ratio H\,{\sc i} H$\alpha$/H$\beta$ = 2.85 \citep{HIcoe} combined with Galactic extinction).  The vertical ranges of the three panels are set to accommodate the intensity of H$\alpha$. 
            \label{fig:ori}} 
            \end{center}
            \end{figure*}

            Since 2018, the MRS-N survey has begun with a resolution of $R\sim$7500, including the blue band ($\sim$4950--5350\,{\AA}) and the red band ($\sim$6300--6800\,\AA) of the spectra. Until now, the MRS-N project has covered about 1300 square degrees in the northern Galactic plane, in the Galactic longitude range $80^\circ < l < 215^\circ$ and the Galactic latitude range $-5^\circ < b < 5^\circ$. Each plane was assigned exposure times 3$\times$900\,s. The survey has used additional plates to cover the field in specific areas selected in the survey plan, such as the one that covers the Rosette Nebula, producing a spatial resolution of $\sim$0.5$^{\prime}\times$0.5$^{\prime}$.  An overview of scientific goals and survey plans of the LAMOST MRS-N project can be found in \citet{mrsn}, \citet{mrsnRV} and \citet{mrsnHa}.  All of the spectra have been processed by the official pipeline of the MRS-N project. The pipeline includes removing cosmic rays, combining data in the same observation, calibrating wavelength, subtracting sky emission, and fitting and measuring emission lines in the spectra. After checking that the emission lines in each spectrum are composed of only one Gaussian element, all of the emission lines are fitted by a single Gaussian function. One-dimensional spectra are shown in Fig \ref{fig:ori}. Detailed processes are described in \cite{mrsnpipline}.

\begin{figure*}[ht!]
            \plotone{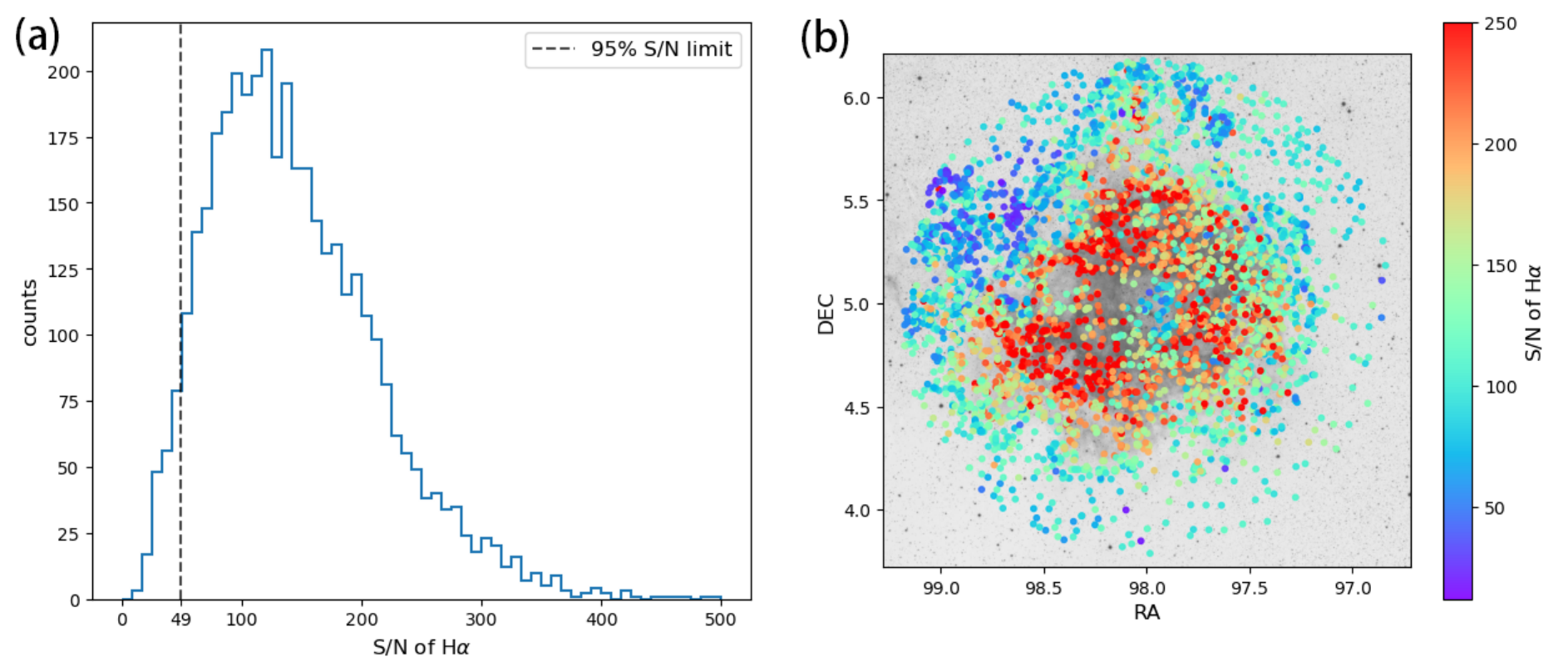}
            \caption{The histogram (panel a) and the spatial distribution (panel b) of the S/N ratio of H$\alpha$ as measured from the fiber spectra of the Rosette Nebula.  Within the optical boundary of the nebula, the S/N's are mostly $>$100. 
            \label{fig:SNH}}
            \end{figure*}

            According to the measure of the H$\alpha$ emission lines in \cite{fountain1979}, in the region where the Monoceros Loop, the supernova remnant in the northeastern direction of the nebula, overlaps with the H\,{\sc ii} region, the line width is larger than in the central part of the Rosette Nebula. Thus, in order to observe whether the SNR has any effect on the gas in the Rosette Nebula, we selected spectra in the field that cover 2 times the radius of the optical boundary of the Rosette Nebula (about $1.2^{\prime}$). To obtain reliable spectra of the nebula, we rejected stellar spectra and invalid spectra. We then checked the original spectra data, deleting spectra with faulty data. After selecting the signal-to-noise ratio of the H$\alpha$, [N\,{\sc ii}], and [S\,{\sc ii}] emission lines, with S/N \textgreater 10, we finally obtained a total of 3854 spectra. Figure \ref{fig:SNH} shows the spatial distribution of the selected spectra. It is evident that the S/N ratio of the observation data is primarily greater than 100 within the nebula's optical boundary. In the area overlapped by the SNR, the S/N ratio of the spectra is relatively lower.

        \subsection{Correction of Line Widths}

            Emission lines can be broadened for various reasons, including natural broadening, energy disturbances, the thermal motion of particles, and instrumental broadening. To subtract instrumental broadening from the observed full-width at half-maximum (FWHM), researchers commonly used skylines in the spectra to estimate instrumental broadening effects. In this work, we used the $\lambda6553$ skyline in each spectrum to correct the FWHM. The correction was made mainly on the basis of Equation \ref{fwhmcor}:

            \begin{equation}\label{fwhmcor}
                \text{FWHM}^{2}_{real} = \text{FWHM}^{2}_{obs} - \text{FWHM}^{2}_{inst},
            \end{equation}

            In this formula, the corrected FWHMs are denoted by FWHM$_{real}$, the observed FWHMs by FWHM$_{obs}$, and the instrumental FWHMs by FWHM$_{inst}$, which is equivalent to the FWHM of the skyline $\lambda6553$ in each spectrum. 

            We first eliminated the spectra using $\text{FWHM}_{obs} - \sigma_{\text{FWHM}obs}$ \textless FWHM$_{inst} + \sigma_{\text{FWHM}inst}$ during the correction procedures. This indicates that the values of FWHM$_{obs}$ are less than those of FWHM$_{inst}$. $\sigma_{\text{FWHM}obs}$ is the uncertainty of FWHM$_{obs}$. For H$\alpha$ lines, 95\% $\sigma_{\text{FWHM}obs}$ are less than 2.21 km s$^{-1}$. For the [N\,{\sc ii}] emission lines, 95\% $\sigma_{\text{FWHM}obs}$ are less than 3.53 km s$^{-1}$. After selecting spectra that satisfy FWHM$_{real} > 3 \sigma_{\text{FWHM}real}$, we finally obtained 3652 spectra for H$\alpha$ lines, 2880 spectra for [N\,{\sc ii}] lines, and 2936 for [S\,{\sc ii}] lines.
            For these spectra, 90\% of the FWHM uncertainties of the H$\alpha$ lines are smaller than 4 km s$^{-1}$, 90\% of those of [N\,{\sc ii}] lines are smaller than 5 km s$^{-1}$, and 90\% of those of the [S\,{\sc ii}] lines are smaller than 3.5 km s$^{-1}$.

            In subsequent sections of this work, the FWHMs that are not annotated refer to the updated FWHMs.

    \section{Emission Line Measurements} \label{sec:results}

        The LAMOST MRS-N project provides us with comprehensive data on the entire nebula. Figure \ref{fig:SNH} has shown the histogram and the spatial distribution of the S/N ratio of 3854 spectra total within the region of the Rosette Nebula. We derive the physical parameters of the H$\alpha$, [N\,{\sc ii}], and [S\,{\sc ii}] emission lines in the region from these spectra with high S/N ratios.

        In the following sections, RVs, FWHMs, and relative intensities of these emission lines are displayed. These physical parameters are first presented as a histogram analysis of the value, which is followed by a scatter plot showing the spatial distribution throughout the area.

        \subsection{Relative Intensities of Emission Lines} \label{sec:I}

            The intensity of the emission lines serves as an indicator of the elemental abundance and energy of the emitted radiation. In this study, we calculated the relative intensities of the H$\alpha$, [N\,{\sc ii}], and [S\,{\sc ii}] emission lines using the $\lambda6553$ skyline in each spectrum, allowing for a comparative analysis of line intensities across the entire field.

\begin{figure*}[ht!]
            \plotone{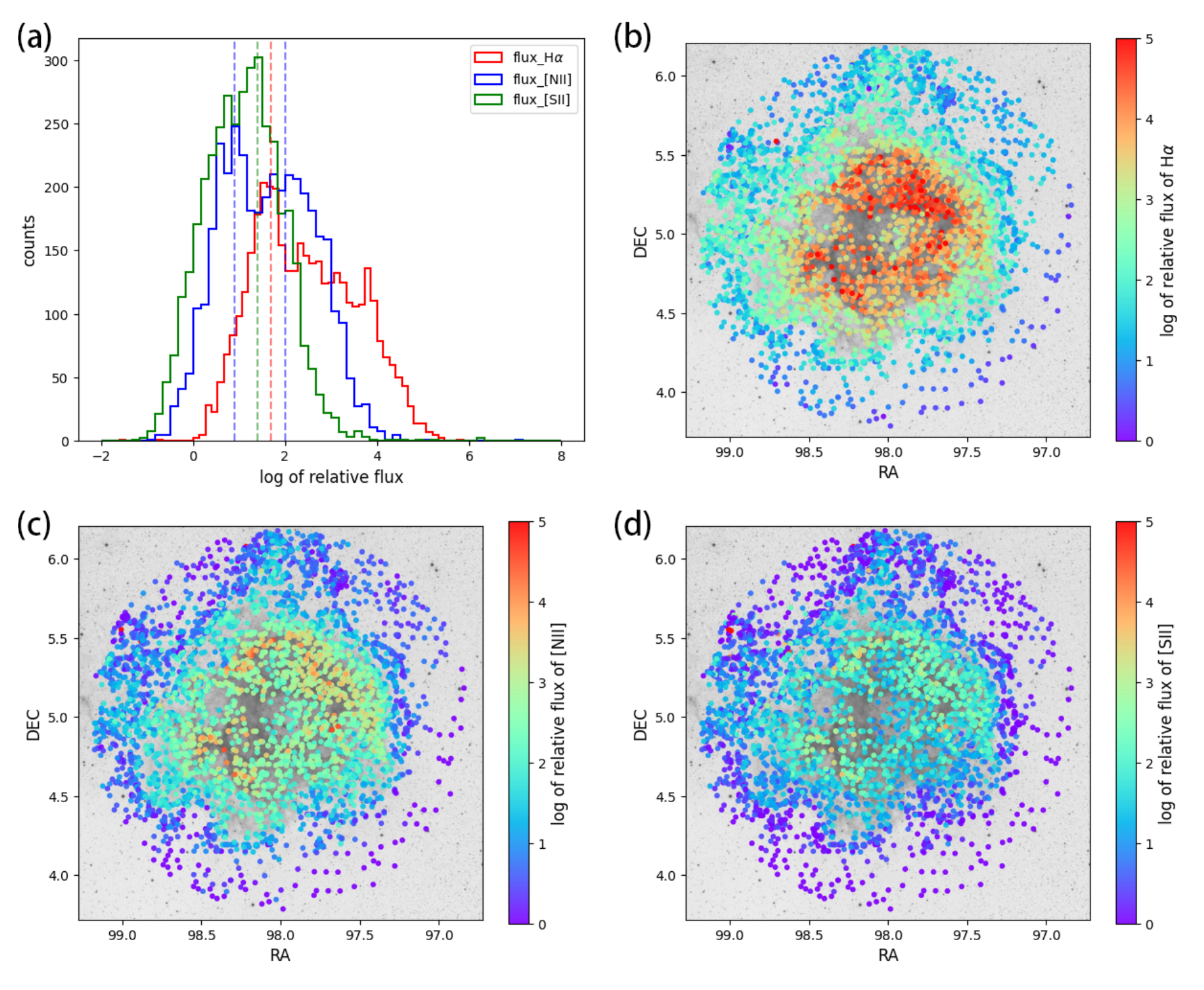}
            \caption{The histograms (panel a) and 2D spatial distributions of the relative intensities of H$\alpha$ (panel b), [N\,{\sc ii}] (panel c), and [S\,{\sc ii}] (panel d).  In panel (a), the number peaks of the distributions of the three emission lines are marked by vertical dashed lines, with the same color-coding as the histograms.  Note that the color scales for panels b, c and d are the same.
            \label{fig:flux}}
            \end{figure*}

            Figure \ref{fig:flux}(a) presents a logarithmic histogram illustrating the relative intensities of the H$\alpha$, [N\,{\sc ii}], and [S\,{\sc ii}] emission lines. The histogram exhibits a unimodal distribution. The H$\alpha$ line demonstrates the highest intensity, characterized by a subtle double-peak structure. The primary peak is observed at a value of 1.7, with a broader peak extending from 2 to 4. The intensity histogram of the [S\,{\sc ii}] emission line reaches its maximum at 1.4, while the intensity of [N\,{\sc ii}] displays a bimodal distribution with peaks at 1 and 2, respectively. 

            Figures \ref{fig:flux}(b), (c) and (d) show the spatial distribution of the relative intensities, with a gradient from weak to strong intensity represented by a color scale ranging from blue to red. The intensities of all three emission lines are predominantly concentrated within the optical boundary of the nebula, characterized by a pronounced edge of intensities along this boundary. To the north of the nebula, a high-intensity structure is observed extending from the boundary, coinciding with the region where the nebula intersects with the Monoceros Loop. In contrast, the central region exhibits diminished intensities, corresponding to the location of the central void within the nebula and the radiation source NGC 2244.

        \subsection{Radial Velocities}\label{sec:RVs}

            In this work, all RVs presented have been adjusted to reflect the local standard of rest (LSR) velocity. This adjustment was achieved by initially converting the relative velocities to heliocentric velocities, followed by a transformation that accounts for the Sun's motion within the galaxy. Consequently, these corrected RVs can be utilized to investigate the dynamics of the gas.

\begin{figure*}[ht!]
            \plotone{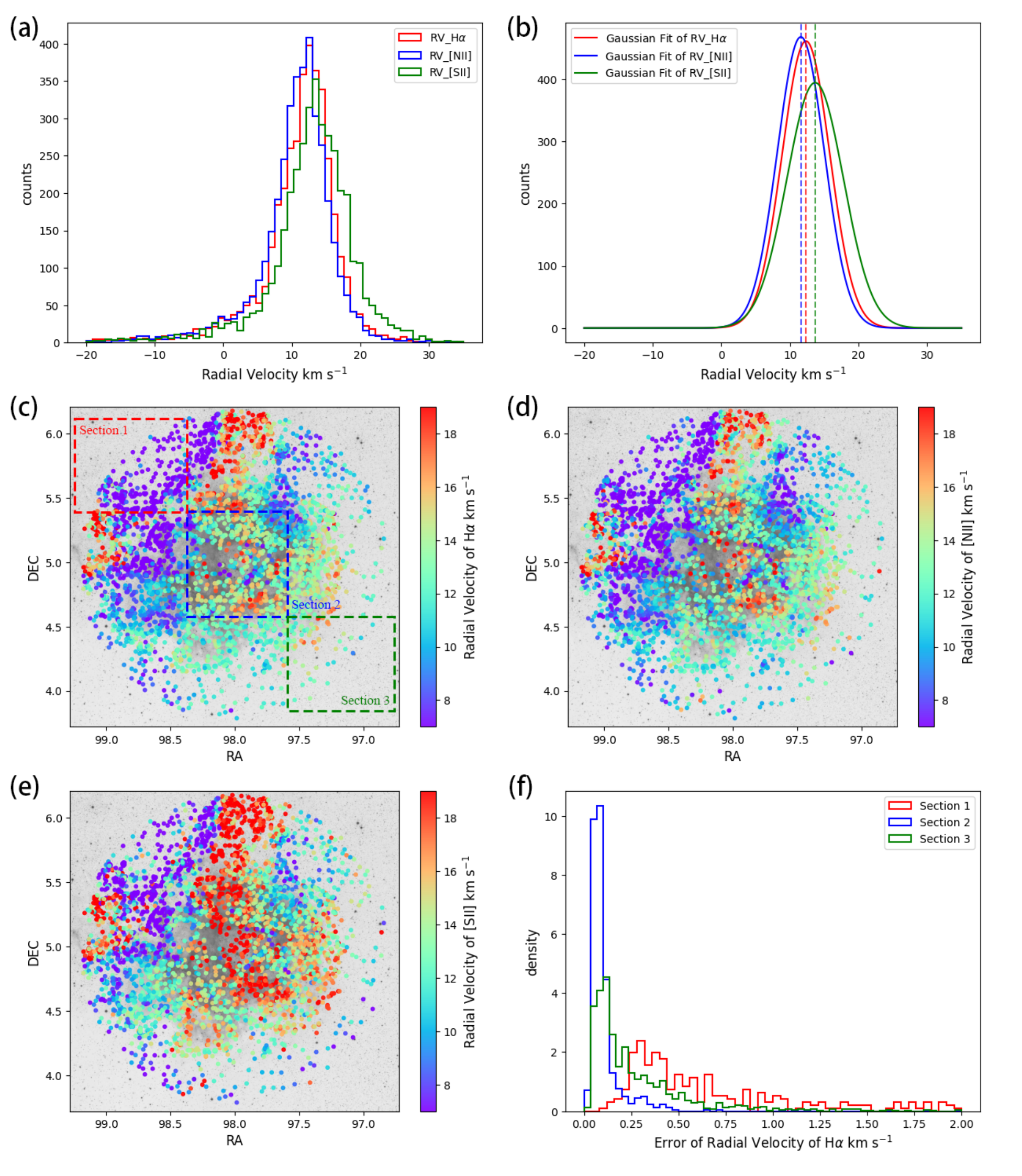}
            \caption{The histograms (panel a) and 2D spatial distributions of the RVs (in km\,s$^{-1}$) measured with H$\alpha$ (panel c), [N\,{\sc ii}] (panel d) and [S\,{\sc ii}] (panel e).  Panel (b) shows the Gaussian-profile fits of the three histograms in panel (a). Panel (f) shows the error of RVs (in km\,s$^{-1}$) measured in three regions shown in panel (c). Notably, the value of error in the northeast region is larger than in other parts of the nebula.
            \label{fig:rv}}
            \end{figure*}

            Figure \ref{fig:rv}(a) presents the histogram of RVs of the emission lines, while Figure \ref{fig:rv}(b) illustrates the Gaussian fits of these histograms. The RVs of the three emission lines exhibit a narrow, single-peaked distribution, while at a value greater than about 17 km s$^{-1}$, the RVs of the [S\,{\sc ii}] emission lines are larger. The RVs of the three lines predominantly fall within the range of 15 to 40 km s$^{-1}$, indicating a pronounced peak distribution. Specifically, the peak RV for the H$\alpha$ emission lines is recorded at 12.33 km s$^{-1}$, for the [N\,{\sc ii}] emission lines at 11.59 km s$^{-1}$, and for the [S\,{\sc ii}] emission lines at 13.67 km s$^{-1}$. The differences between these peak values are approximately 1 km s$^{-1}$.

            Figures \ref{fig:rv} (c), (d) and (e) illustrate the spatial distribution of the RVs of the three emission lines. The RV values are represented using a color gradient, ranging from blue for lower values to red for higher values, with consistent color scales applied across all three scatter plots. Analysis of these scatter maps reveals a prominent high-velocity, bow-shaped feature situated in the central region of the nebula, observable in the RVs of all three emission lines. This feature exhibits velocities that exceed 16 km s$^{-1}$, which is approximately 4 km s$^{-1}$ higher than those in other regions of the nebula. The structure is the most clear in the spatial distribution of the RVs of the [S\,{\sc ii}] emission lines.

            Figure \ref{fig:rv} (f) shows the spatial distribution of the uncertainties of the RVs. The three histograms display the errors in RVs derived from H$\alpha$ emission lines within the three distinct regions depicted in panel (c), each region being indicated by a different color. It is evident that the uncertainties in the RVs are larger in the northeastern region of the nebula compared to the central and southwestern regions. This increased uncertainty is likely attributable to the influence of the background Monoceros Loop.

            \cite{RVsample} used stigmatic grating spectroscopy to determine the heliocentric velocities of the Rosette Nebula, yielding a measurement of 32.8$\pm$0.5 km s$^{-1}$.  In contrast, our analysis indicates that the heliocentric radial velocity (RV) peaks at approximately 30 km s$^{-1}$, exhibiting considerable variability among individual data points. This finding is generally consistent with the results reported by \cite{RVsample}. Furthermore, the LSR velocity aligns with the value reported by \cite{thesis2017}, approximately 26 km s$^{-1}$. Notably, the bow-shaped structure observed in our data was not identified in the work of \cite{RVsample}, likely due to the limited spatial resolution of their spectral data.

            In the northeastern region of the figures, where the field intersects the supernova remnant, the RVs are observed to be lower than those in other areas of the nebula, with a value of less than 10 km s$^{-1}$.

\begin{figure*}[ht!]
            \plotone{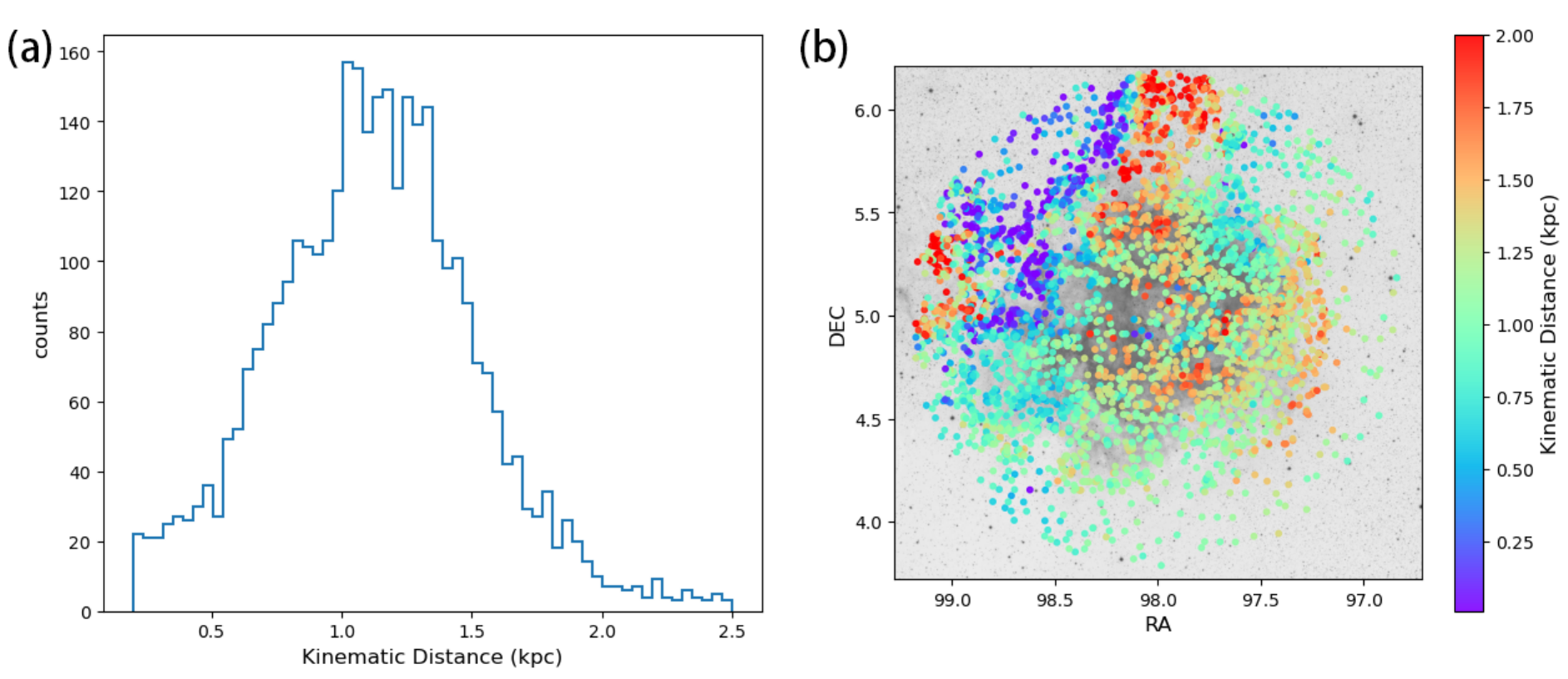}
            \caption{The histogram (panel a) and 2D spatial distribution (panel b) of the kinematic distance (in unit of kpc) measured with the H$\alpha$ emission line.
            \label{fig:kdall}}
            \end{figure*}

            In addition to the spatial distribution of the RVs, we also used the RVs of the H$\alpha$ emission lines to estimate the kinematic distance of the nebula. This estimation was performed using the Monte Carlo method, as detailed in \cite{kdMC}. The findings are illustrated in Figure \ref{fig:kdall}. The scatter plot indicates that within the optical boundary of the Rosette Nebula, the estimated kinematic distance is approximately 1.5 kpc, which aligns with the results in \cite{zhao2018}. Moreover, within the area where the Monoceros Loop overlaps with the nebula, the kinematic distance measurement is significantly reduced, a phenomenon that is probably attributable to the substantial uncertainty associated with the RV values.

        \subsection{Line Widths}

\begin{figure*}[ht!] 
            \plotone{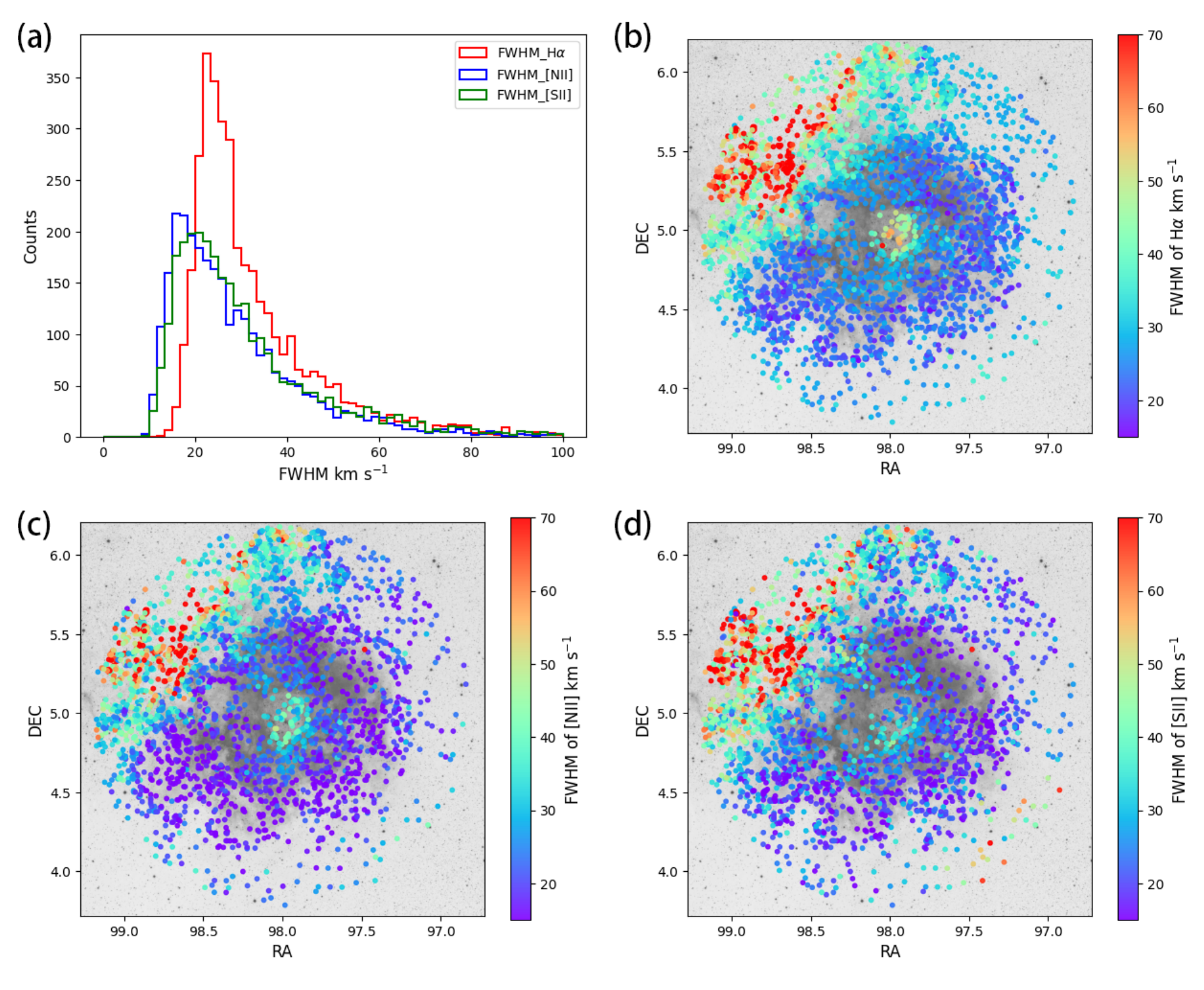}
            \caption{The histograms (panel a) and the 2D spatial distributions of the FWHMs of H$\alpha$ (panel b), [N\,{\sc ii}] (panel c) and [S\,{\sc ii}] (panel d). The color scales for panels b, c and d are the same.
            \label{fig:fwhm}}
            \end{figure*}

            The widths of emission lines, the corrected FWHMs, provide insight into various parameters, including temperature and gas pressure. As illustrated in Figure \ref{fig:fwhm}(a), the histogram depicting the FWHMs of the three emission lines reveals single-peak distributions characterized by a considerable degree of dispersion. Taking into account the uncertainties of the [N\,{\sc ii}] and [S\,{\sc ii}] emission lines, the FWHM distributions of the three emission lines exhibit a general consistency, predominantly falling within the range of 15 to 60 km s$^{-1}$, with a peak occurring at approximately 23 km s$^{-1}$. The value generally agrees with results in \cite{RVsample}.

            Figures \ref{fig:fwhm} (b), (c) and (d) illustrate the spatial distribution of FWHMs. Within the areas covered by the nebula, the FWHMs of the three emission lines are relatively narrow, predominantly falling within the range of 20 to 30 km s$^{-1}$. In the northeast section of the nebula, which intersects the SNR region, the FWHMs exhibit broader values, spanning 30 to 40 km s$^{-1}$. In the central region, where the NGC 2244 cluster is located, the FWHMs are considerably wider than the surroundings, exceeding 40 km s$^{-1}$. Excluding the central and northeast regions of the nebula, the average FWHM of the H$\alpha$ emission line is approximately 25 km s$^{-1}$, aligning with the findings reported in \cite{RVsample}. The observed distributions are consistent with those of the relative intensities.

            Using FWHM data of emission lines, it is possible to approximate the temperature of the nebula. \cite{FWHMtemp} identified a correlation between the FWHMs of the H$\alpha$ and [N\,{\sc ii}] emission lines and the gas temperature $T$, as well as the turbulent velocity $V_{t}$. This relationship is articulated in the subsequent equations.

            \begin{equation}
                T(\text{K}) = 23.5\, W_{\text{H}}^{2}\,\left(1\,-\,\frac{W_{\text{N}}^{2}}{W_{\text{H}}^{2}}\right),
            \end{equation}\\
            and
            \begin{equation}
                V_{t}(\text{km\,s}^{-1}) = 1.04\,W_{\text{N}}\,\left(1\,-\,0.071\,\frac{W_{\text{H}}^{2}}{W_{\text{N}}^{2}}\right)^{1/2},
            \end{equation}\\

\begin{figure*}[ht!]
            \plotone{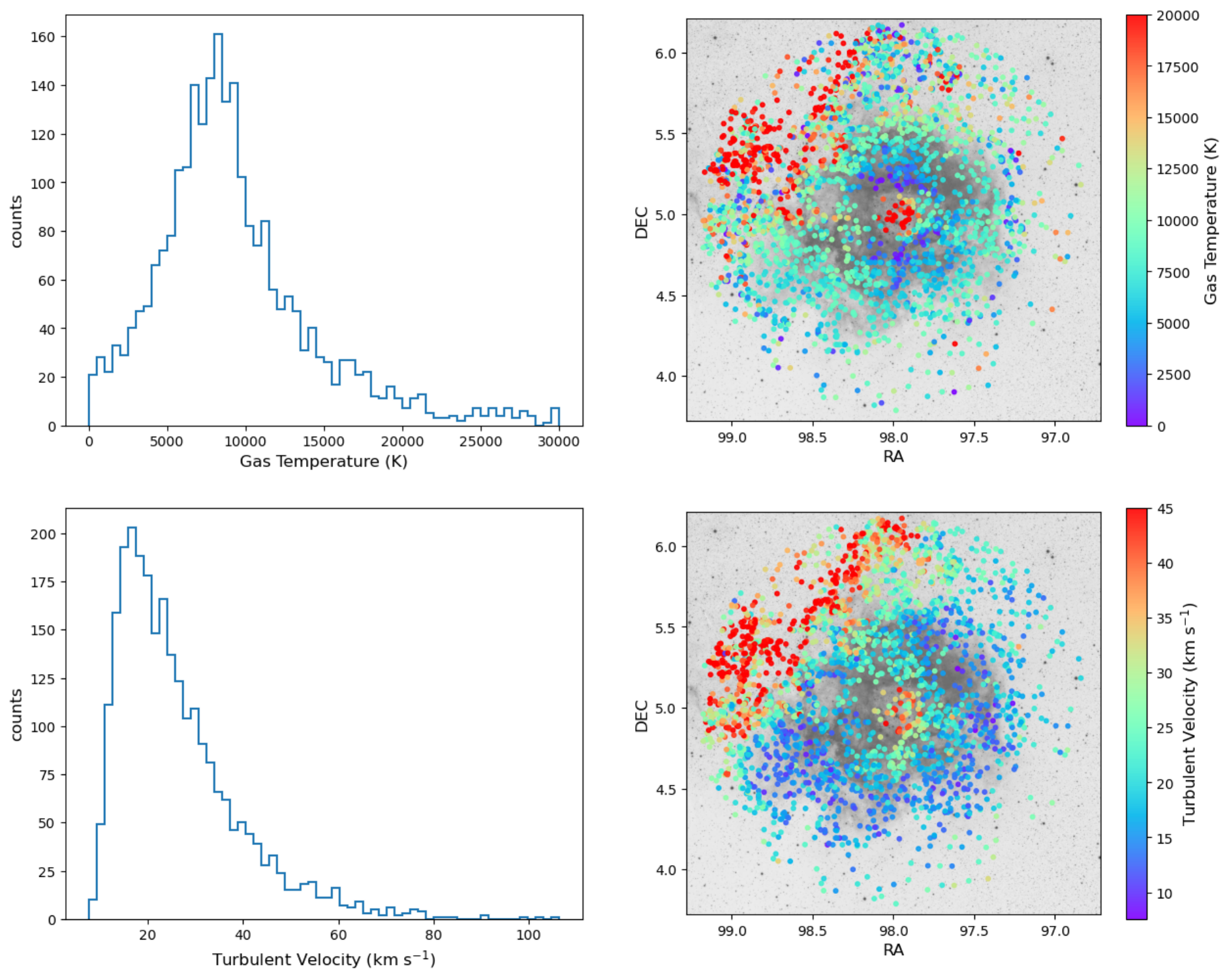}
            \caption{The gas temperatures (top panels) and turbulent velocities (bottom panels) of the Rosette Nebula.  Panel (a) and (b) show the histogram and the 2D spatial distribution of gas temperature (in K). Panel (c) and (d) show the histogram and the spatial distribution of turbulent velocity.
            \label{fig:tv}}
            \end{figure*}

            In the equations presented, $W_{\text{H}}$ and $W_{\text{N}}$ denote the FWHMs of the emission lines H$\alpha$ and [N\,{\sc ii}] in the same volume of gas, respectively. Figures \ref{fig:tv}(a) and (b) represent the histograms and spatial distribution of the gas temperature, revealing a unimodal distribution that peaks at approximately 8000 K. Figures \ref{fig:tv} (c) and (d) depict the histogram and spatial distribution of the turbulent velocity of the gas. Within the optical boundary of the nebula, the turbulent velocity predominantly ranges from 10 to 40 km s$^{-1}$. In the northeastern region, where the SNR is situated, both the gas temperature and the turbulent velocity are higher compared to other parts depicted in the figures. In addition, it is observed that the spatial distribution of the FWHM does not align with the gas temperature; specifically, the gas temperature is higher at the nebula's periphery than in its interior, whereas the FWHMs are comparatively smaller.

        \subsection{The Parameter Table}

            In order to consolidate the physical parameters obtained, we have compiled parameter tables of each emission line to present the findings, and these tables are ready to be published. The parameter tables provided in this study include the coordinates of the data points in (R.A., Dec.), RVs of each emission line in km s$^{-1}$, FWHMs along with uncertainties of each emission line in km s$^{-1}$, as well as relative intensities of the three emission lines. Parameter tables are partly displayed as shown in Table \ref{tab:H}.

            \begin{deluxetable*}{ccccccccccc}
            \tabletypesize{\scriptsize}
            \tablewidth{0pt} 
            \tablecaption{Nebular Parameters Measured Using the H$\alpha$ Emission Line in the LAMOST MRS-N Spectra \label{tab:H}}
            \tablehead{
            \colhead{R.A.}& \colhead{Dec.} & \colhead{ID} &
            \colhead{SN} & \colhead{RV} & \colhead{err\_RV} & \colhead{FWHM} & \colhead{err\_FWHM} &
            \colhead{relative flux} & \colhead{err\_relative flux}\\
            \colhead{(J2000)} & \colhead{(J2000)} & \colhead{} & \colhead{} &
            \colhead{($\rm km~s^{-1}$)} & \colhead{($\rm km~s^{-1}$)} &
            \colhead{($\rm km~s^{-1}$)} & \colhead{($\rm km~s^{-1}$)} &
            \colhead{(relative to $I_{\lambda6553}$)} & \colhead{} 
            } 
            \startdata 
                    99.01606 & 4.8689737 & MN063604+045208N & 91 & 9.59 & 0.37 & 42.6 & 1.9 & 3.79 & 0.19  \\ 
                    98.924324 & 4.6189737 & MN063542+043708N & 169 & 10.22 & 0.1 & 26.6 & 1.6 & 5.23 & 0.21  \\ 
                    99.116394 & 4.9023075 & MN063628+045408N & 61 & 14.23 & 0.42 & 39.6 & 1.8 & 5.44 & 0.25  \\ 
                    99.149925 & 4.802307 & MN063636+044808N & 108 & 12.78 & 0.29 & 32.5 & 1.8 & 4.12 & 0.18  \\ 
                    98.89056 & 4.9106407 & MN063534+045438N & 118 & -0.37 & 0.21 & 42.8 & 1.6 & 10.29 & 0.64  \\ 
                    99.03276 & 4.893974 & MN063608+045338N & 121 & 2.65 & 0.23 & 41 & 2.1 & 7.92 & 0.58  \\ 
                    98.05425 & 4.8689737 & MN063213+045208N & 172 & 10.59 & 0.1 & 33.3 & 2.3 & 39.84 & 2.58  \\ 
                    98.1464 & 4.802307 & MN063235+044808N & 296 & 16.29 & 0.05 & 24.1 & 4.3 & 79.22 & 7.12  \\ 
                    98.12148 & 4.7273073 & MN063229+044338N & 276 & 14.56 & 0.05 & 27.9 & 2.3 & 70.57 & 4.69  \\ 
                    98.17188 & 4.6189737 & MN063241+043708N & 248 & 7.49 & 0.05 & 32.5 & 2.9 & 137.78 & 18.19 
            \enddata
            \tablecomments{Table \ref{tab:H} is published in its entirety in the machine-readable format in the online Journal. Included with this version is a single table merging the H$\alpha$, [N\,{\sc ii}] and [S\,{\sc ii}]. A portion is shown here for guidance regarding its form and content.}
            \end{deluxetable*}

    \section{Results and Discussion} \label{sec:discussion}

        \subsection{Uncertainties in Electron Density}

            The electron density can be determined by the ratio of intensities of the [S\,{\sc ii}] lines. As stated by \cite{electrondensity}, for an electron temperature $T_{e}$ = 10000 K, let $R = \frac{I_{6716}}{I_{6732}}$, where $I_{6716}$ and $I_{6732}$ represent the intensities of forbidden lines $\lambda6716$\AA, and $\lambda6732$\AA, respectively. Consequently, the following equation can be derived.
            
            \begin{equation}\label{eq:elec}
            \begin{split}
            \log(n_{e}[ \text{cm}^{-3}]) = 0.0543\tan (-3.0553R +2.8506)\\ + 6.98 -10.6905R+9.9186R^{2}-3.5443R^{3},
            \end{split}
            \end{equation}

            This equation is applicable for $R$ \textless 1.42. From this we derive an average electron density of 39.45 cm$^{-3}$ within the Rosette Nebula based on available data points. Given that the diameter of the Rosette Nebula is approximately 30 pc, as noted by \cite{nebook}, the Rosette Nebula may be classified as a giant H\,{\sc ii} region rather than a classical H\,{\sc ii} region, as previously understood.

            In \cite{radiopara}, the average electron density value was calculated to be 16.2$\sim$18.4 cm$^{-3}$ from radio observation data, a value slightly lower than those reported in the current study, but still of the same magnitude. The discrepancy may be attributed to the differing methodologies employed in obtaining the results, as radio observations primarily focus on optically thin gas. In addition, the assumption of constant temperature and density within the H\,{\sc ii} region in \cite{radiopara} could also influence the estimated electron density. In this work, certain [S\,{\sc ii}] intensity ratios in this study exceed the valid range defined by Eq.\ref{eq:elec}, which may contribute to uncertainty in the findings.

            \subsection{Distribution of Chemical Abundances} \label{sec:chemical}

            Various methods have been used to analyze the chemical abundance of interstellar gas. The predominant approach for investigating chemical abundance involves the analysis of the intensities of emission lines that are sensitive to metallicity. Numerous models have been developed to quantify abundance in H\,{\sc ii} regions. Given the constraints posed by the limited spectral data available at the red end, this study used the $N2S2H$ parameter to determine the chemical abundance of the nebula.

            Researchers commonly employ the $N2$ parameter, defined as $\frac{I_{\lambda6584}}{I_{\text{H}\alpha}}$, for the assessment of the chemical abundance. According to \cite{N2prob}, the linear relationship between $\log\frac{I_{\lambda6584}}{I_{\text{H}\alpha}}$ and $12 + \log \frac{[\text{O\,III}]}{\text{H}\beta}$ exhibits variability, which may result in an underestimation of the chemical abundance. In addition, this model is significantly influenced by the ionization fraction of the gas. Consequently, in this work, we choose to use the $N2S2H$ parameter.

            The $N2S2H$ parameter encompasses the intensities of three emission lines, including the [N\,{\sc ii}] $\lambda6584$\AA, [S\,{\sc ii}] $\lambda6716$\AA\, and [S\,{\sc ii}] $\lambda6731$\AA\, forbidden lines and the H$\alpha$ emission line.  A theoretical calibration of this model has been previously established by \cite{chemical}; however, it remains sensitive to the relationship between the nitrogen-to-oxygen (N/O) and oxygen-to-hydrogen (O/H) ratios.  The definition of the parameter used in this model is provided as 

            \begin{equation}
                N2S2H = \log\frac{I_{\lambda6584}}{I_{\lambda6716} + I_{\lambda6731}} + 0.264\log \frac{I_{\lambda6584}}{I_{\text{H}\alpha}},
            \end{equation}
            and 
            \begin{equation}
                12 + \log \frac{[\text{O\,III}]}{\text{H}\beta} = 8.77 + N2S2H.
            \end{equation}
            The equations demonstrate strong linearity when the [O\,{\sc iii}]/H$\beta$ line ratio $12 + \log \frac{[\text{O\,III]}}{\text{H}\beta} < 9.05$ is satisfied. As noted in \cite{chemical}, this method is deemed reliable, exhibiting minimal residual dependence on additional physical parameters. Thus, in this work, we used the result of $12 + \log \frac{[\text{O\,III]}}{\text{H}\beta}$ to analyze the chemical abundance of the nebula.

\begin{figure*}[ht!]
            \plotone{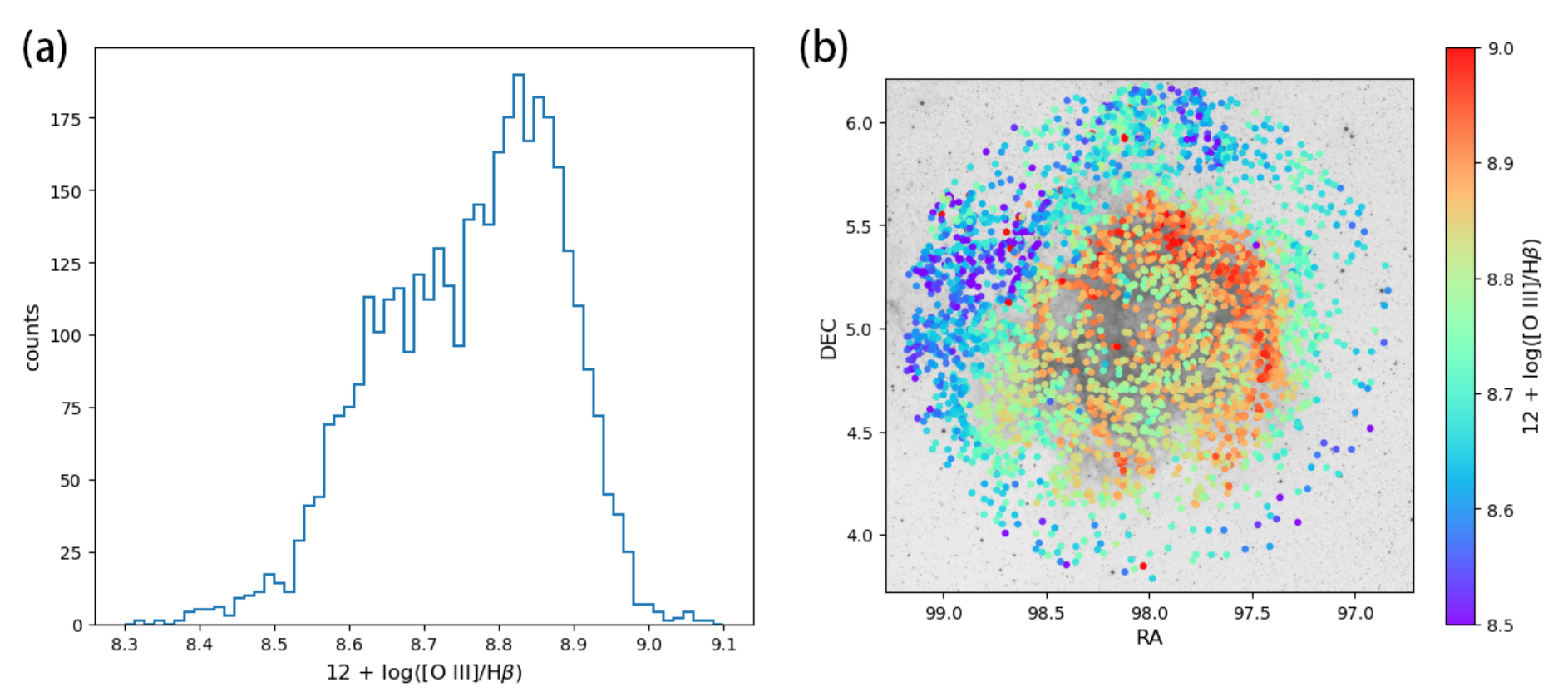}
            \caption{The histogram (panel a) and the spatial distribution (panel b) of 12+$\log{[{\rm O\,III}]/{\rm H}\beta}$, which is used as a proxy of chemical abundance. 
            \label{fig:chem}}
            \end{figure*}

            Figure \ref{fig:chem}(a) presents a histogram depicting the distribution of the chemical abundance, characterized by a bimodal structure. Figure \ref{fig:chem} (b) shows the spatial distribution of the chemical abundance. The scatter plot indicates that the chemical abundance is elevated in the outer regions of the nebula, exhibiting a pronounced edge at the optical boundary. Outside this boundary, the chemical abundance is relatively low, with even lower values observed in the direction of the SNR. The average chemical abundance within the optical boundary of the nebula is $12 + \log \frac{[\text{O\,III}]}{\text{H}\beta} = 8.859$.

            The spatial distribution of the chemical abundance of the HII region represents the chemical environment of the Galaxy and provides insights into the most recent generation of stars. A distinct circular formation of elevated chemical abundance surrounds the central cluster NGC 2244, implying that this distribution pattern may result from historical star-formation activities within the OB cluster. During these events, high-mass stars expelled gas enriched with higher chemical abundances. Furthermore, ongoing star formation processes are observed in structures such as ``elephant trunks" and ``globulettes" within the nebula, which also contribute to the enhancement of chemical abundance \citep{globulettes}.

\begin{figure}[ht!]
            \plotone{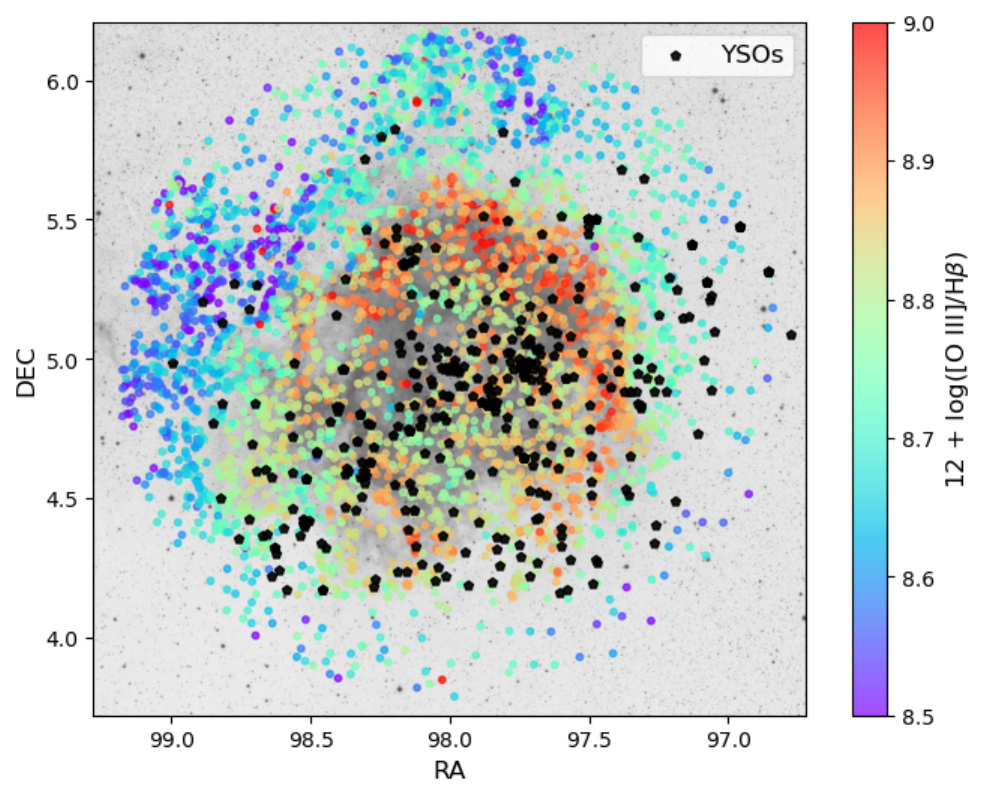}
            \caption{The spatial distribution of 12+$\log{[{\rm O\,III}]/{\rm H}\beta}$ of the Rosette Nebula, overlaid with the locations of YSOs adopted from the literature (see text in Section\,4.2).  Significant concentration of YSOs is noticeable in the central region of the nebula.  However, the distances to these YSOs are still not yet clear.
            \label{fig:chemYSO}}
            \end{figure}

            In addition to the central cluster, young stellar objects (YSOs) may also influence the distribution of chemical abundance. Evidence from sky surveys and observations of YSOs indicates that star formation processes are ongoing within the nebula's vicinity and that these processes interact with one another \citep{YSO1, YSO2, YSO3, YSO4, YSO5, YSO6, YSO7, YSO8, YSO9}. Figure \ref{fig:chemYSO} depicts the spatial relationship between the YSOs acquired from these catalogs and the chemical abundance values. The location of YSOs concentrates on the central region of the nebula, where the chemical abundance is relatively low. This observed correlation suggests a potential association between the YSOs and the areas with reduced chemical abundance.  However, as little distance information was provided in the catalog, the association remains untrustworthy. Further study about distances could clarify the assumption.

        \subsection{Surrounding DIG and ionization fraction}\label{sec:DIG}

\begin{figure*}[ht!]
            \plotone{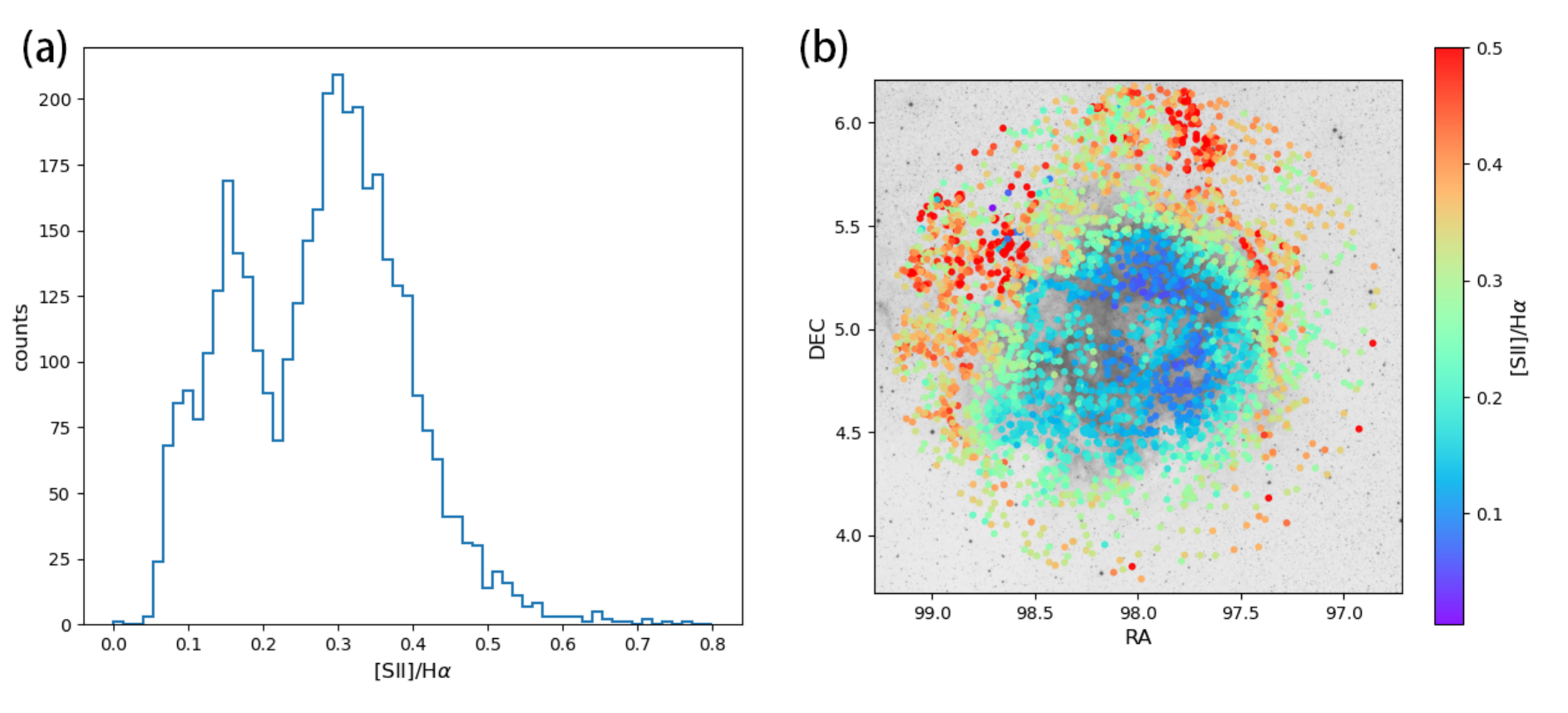}
            \caption{The histogram (panel a) and the spatial distribution (panel b) of [S\,{\sc ii}]/H$\alpha$ emission line intensities. Within the optical boundary of the nebula, the values of [S\,{\sc ii}]/H$\alpha$ are smaller than 0.3.
            \\
            \label{fig:StoH}}
            \end{figure*}

\begin{figure*}[ht!]
            \plotone{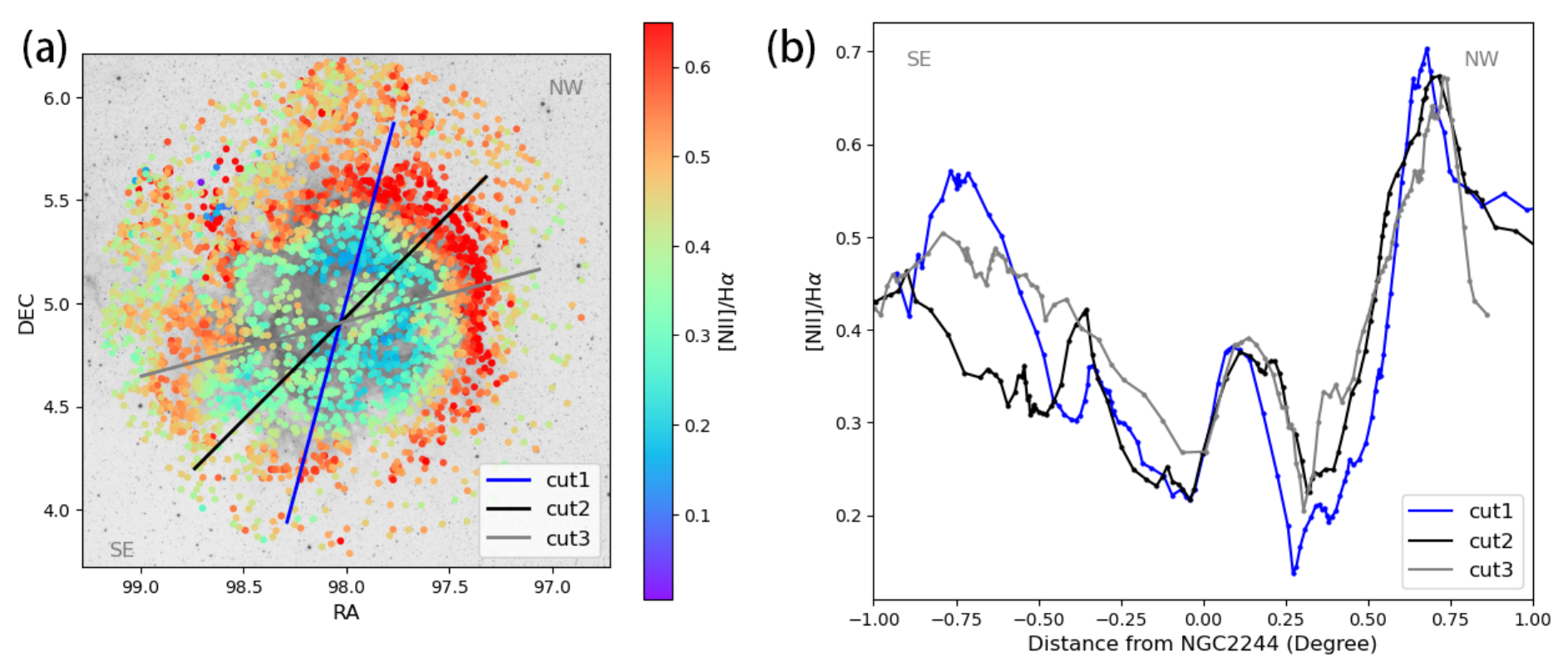}
            \caption{Panel (a) illustrates the spatial distribution of the ratio of the [N\,{\sc ii}]/H$\alpha$ emission line intensities. A pronounced increase in the line ratios is observed in the northwestern regions. Panel (b) presents the intensity ratios along the three designated cuts depicted in panel (a), in which the zero degree matches the center of the OB cluster NGC2244. Notably, the increase in the intensity ratio in the northwestern direction occurs at a consistent distance from the central cluster. 
            \\
            \label{fig:NtoH}}
            \end{figure*}

            The intensity of the H$\alpha$ emission line serves as an indicator of the photonization process induced by ultraviolet photons within H\,{\sc ii} regions, while the intensity of the [S\,{\sc ii}] forbidden line is typically associated with shock waves and collisions. Thus, researchers commonly use the [S\,{\sc ii}]/H$\alpha$ intensity ratio to determine the underlying physical mechanisms responsible for the observed emission lines. This analytical approach was initially introduced by \cite{SIItoHaori}. As illustrated in Figure \ref{fig:StoH}, the [S\,{\sc ii}]/H$\alpha$ intensity ratio within the Rosette Nebula ranges from 0.05 to 0.3, suggesting that the region is predominantly photoionized. At the periphery of the nebula, particularly along the northwestern edge, the [S\,{\sc ii}]/H$\alpha$ ratio exhibits a marginally higher value. In the northeast region of the observational data, which exhibits a notably high [S\,{\sc ii}]/H$\alpha$ intensity ratio and especially high ionization fraction, it is likely that these characteristics are the result of both radiation and collisional effects stemming from the SNR. As illustrated in Figure \ref{fig:NtoH}, the [N\,{\sc ii}]/H$\alpha$ ratio also exhibits a sharp gradient along the edge.


            In H\,{\sc ii} regions, the presence of high-energy radiation leads to the ionization of the surrounding gas, which in turn produces emission lines from a variety of elements. However, the energy of the radiation is not uniform throughout the nebula, leading to variations in ionization levels across different regions. By examining the intensities of the emission lines across the nebular field, it is possible to estimate the ionization fraction of the gas and investigate its influence on other physical parameters.

            The ionization fraction of the gas can be estimated by the ratio of S$^{+}$/S. As indicated by \cite{madsen2006}, this ratio can be derived from the intensity ratio of the [S\,{\sc ii}] and [N\,{\sc ii}] emission lines, using the following equation:

            \begin{equation}
                \frac{[\text{S\,II}]}{\text{[N\,II]}} = 4.62\text{e}^{0.04/T_{4}}\left(\frac{\text{S}^{+}}{\text{S}}\right)\
                \left(\frac{\text{S}}{\text{H}}\right)\left[\left(\frac{\text{N}^{+}}{\text{N}}\right)\left(\frac{\text{N}}{\text{H}}\right)\right]^{-1},
            \end{equation}

            In this work, we assume that the ratios of ionized to neutral species are H$^{+}$/H = 1 and N$^{+}$/N = 0.8. Regarding the metal abundance of the gas, we assume that N/H = 7.5 $\times$ 10$^{-5}$ \citep{NtoHvalue} and S/H = 1.86 $\times$ 10$^{-5}$ \citep{StoHvalue}.

\begin{figure*}
            \plotone{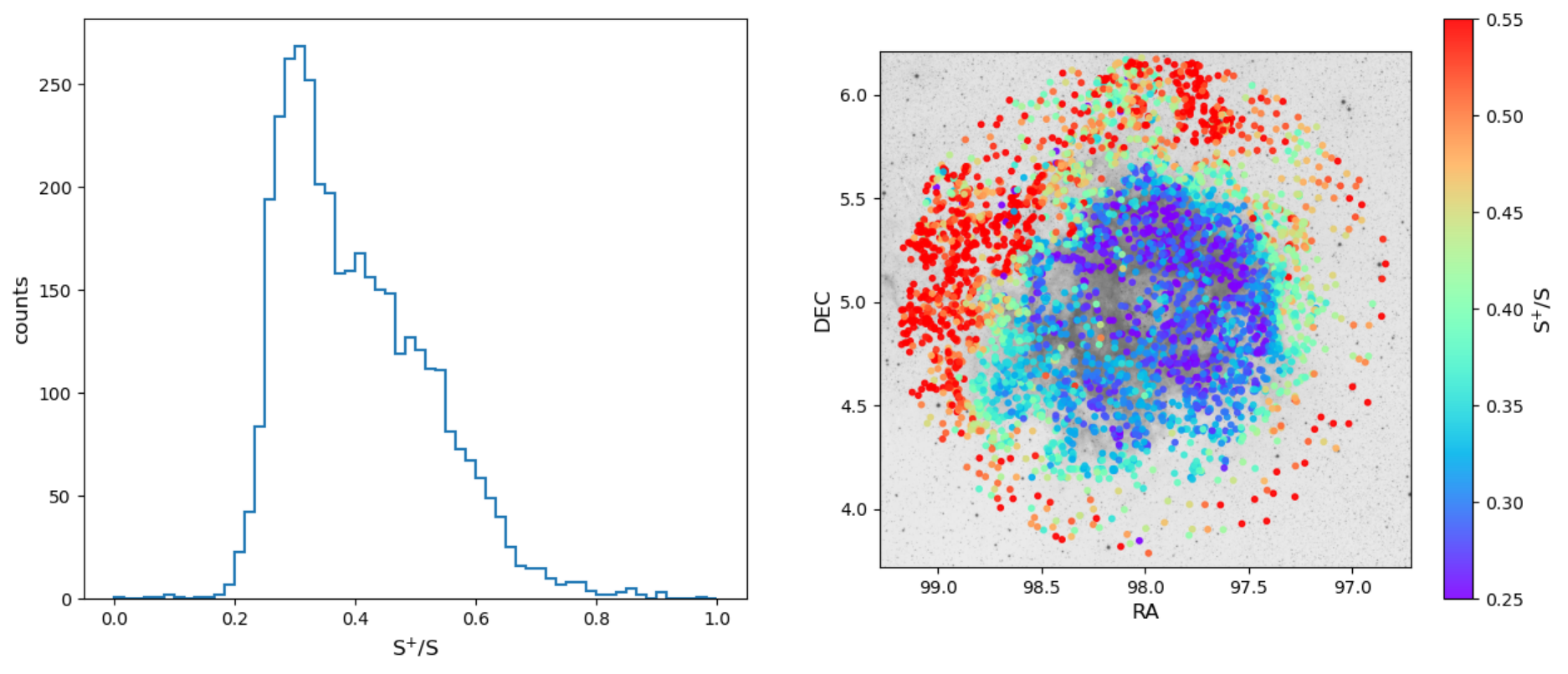}
            \caption{The histogram (panel a) and spatial distribution (panel b) of ionization fraction measured with S$^{+}$/S ratio.
            \label{fig:if}}
            \end{figure*}

            The results are depicted in Figure \ref{fig:if}. It is apparent that within the optical boundaries of the Rosette Nebula, the S$^{+}$/S ratio is markedly lower than that of the surrounding regions, suggesting that sulfur within the nebula exists in a higher ionization state. Conversely, the gas in the vicinity of the nebula exhibits a lower degree of ionization. Within the nebula, the S$^{+}$/S ratio is found to be lower in the central region and increases towards the periphery. Furthermore, the spatial distribution of the ionization fraction aligns with the findings presented in \cite{thesis2017}, which uses the ratios of [O\,{\sc iii}] and [S\,{\sc ii}] emission lines to derive the parameter and similarly indicates an increasing trend towards the edges of the nebula.

            The area adjacent to the H\,{\sc ii} region is generally distinguished by the presence of diffuse ionized gas (DIG). The ratios of [S\,{\sc ii}]/H$\alpha$ and [N\,{\sc ii}]/H$\alpha$ are found to be elevated beyond the optical boundary of the nebula, which is indicative of the characteristics associated with DIG \citep{WIM}. Consequently, it can be inferred that the surrounding region comprises a distinct gas component that differs from that of the H\,{\sc ii} region.

        \subsection{Structures in RV}

            As discussed in Section \ref{sec:RVs}, a bow-shaped high-velocity structure has been identified extending outward from the nebula, in addition to the low-velocity region in the northeastern area where potential interaction with the Monoceros Loop may occur.

            According to \cite{zhao2018}, the Monoceros Loop is located at a distance of 1.98 kpc, resulting in a separation of 0.4 kpc between the supernova remnant (SNR) and the Rosette Nebula. This proximity suggests a low likelihood of direct interaction between the two nebulae; however, the influence of propagating shock waves may still be present, as indicated by various parameters previously discussed.

            Observations presented in Section \ref{sec:DIG} reveal that the gas situated beyond the optical boundary of the nebula is linked to a distinct component. If the high-velocity structure were to be attributed to a foreground object, it would pose challenges in maintaining consistency in the relative fluxes and the widths of the emission lines.

    \section{Summary} \label{sec:summary}

        We present, for the first time, a table of nebular parameters for the Rosette Nebula as derived based on the comprehensive measurements of the multi-fiber, medium-resolution spectroscopy that encompasses the entire nebula.  These new observations were conducted as part of the LAMOST MRS-N project that surveys the nebulae, including H\,{\sc ii} regions, in the northern Galactic plane ($80^\circ < l < 215^\circ$, $-5^\circ <b< 5^\circ$).  For this study, we selected and analyzed 3854 high-quality fiber spectra that cover the whole nebular region of Rosette, including the periphery regions. 

        The nebular parameters reported here include the relative intensities, RVs, and FWHMs as measured using three typical nebular emission lines (H$\alpha$, [N\,{\sc ii}] and [S\,{\sc ii}]) in the red wavelength of the optical region.  The inclusion of extensive parameters that cover the entire Rosette Nebula is crucial for enhancing our understanding of the star formation processes occurring within the nebula, as well as the interactions between the H\,{\sc ii} region and the surrounding ISM, including the SNR Monoceros Loop.  Using the LAMOST MRS-N spectra, we have developed a spectral sample of the Rosette Nebula and its surrounding regions. Our findings of this study are presented below.

        \begin{enumerate}
            \item In the 4.52 square-degree field, a total of 3,854 nebular spectra have been acquired with high S/N. The relative intensities, RVs, and FWHMs of the H$\alpha$, [N\,{\sc ii}] and [S\,{\sc ii}] emission lines have been quantified in order to compile a comprehensive dataset of spectral information of the Rosette Nebula.
            \item The spatial distribution of relative intensities indicates that the emission lines are strong in the bright nebula regions, whereas they exhibit significantly lower intensity in the central cavity and its adjacent areas.  Among the emission lines, the relative intensity of the H$\alpha$ line is the most prominent, followed by the [N\,{\sc ii}] lines, while the [S\,{\sc ii}] lines display the weakest intensity.
            \item The spatial distribution of the RVs reveals distinct characteristics within the luminous nebula. The observed values of RVs indicate that the nebula is receding from the observer, with velocities predominantly ranging from 5 to 15 km s$^{-1}$. In the central region of the field, a bow-shaped area with a relatively high velocity is evident across all three emission lines, with speeds exceeding 16 km s$^{-1}$. This structure could be related to the formation of the nebula.
            \item The spatial distribution of the FWHMs demonstrates uniformity within the optical boundary of the nebula, showing values less than 30 km s$^{-1}$. In contrast, within the central cavity and extending towards the northeast region of the nebula, the FWHM values are observed to be greater than those found in the surrounding areas.
            \item The average electron density, derived from the intensities of the [S\,{\sc ii}] emission lines, is measured at 39.45 cm$^{-3}$. This value exceeds previous estimates obtained from radio observations and deviates from the characteristics typically associated with classical H\,{\sc ii} regions.
            \item The chemical abundance, as determined by the $N2S2H$ parameter, is notably elevated, exhibiting a distinct gradient along the optical boundary of the nebula. The processes of star formation occurring within the central cluster NGC 2244 and in substructures within the nebula are likely contributing factors to the observed increase in the chemical abundance of the gas.
            \item The intensity ratios of the [S\,{\sc ii}]/H$\alpha$ and the [N\,{\sc ii}]/H$\alpha$ emission lines exhibit significant variation along the periphery of the nebula, particularly in the northwestern region. A pronounced gradient is observed at approximately the same distance from the central cluster NGC2244.
        \end{enumerate}

        The parameter Table\,\ref{tab:H} of the Rosette Nebula provides valuable observational data for the community of Galactic nebulae, and will help to lay down the theoretical framework for future research endeavors, particularly in the construction of detailed photoionization models for this H\,{\sc ii} region that take into account the kinematics within.  Moving forward, these parameter tables will facilitate more in-depth observations, including the high-resolution spectroscopy using the integral-field units (IFUs) on large telescopes that allows both chemical and kinematical analysis of the Rosette Nebula with very high precision. One of the best choices of such observations will be using the MEGARA\footnote{\url{https://www.gtc.iac.es/instruments/megara/}} spectrograph on the 10 m Gran Telescopio Canarias (GTC); this has been scheduled in our future plan, which aims to develop benchmark IFU observations of the Rosette Nebula.

        At the time this article is ready for submission after revision, there is another work utilized the data from the SDSS-LVM survey to investigate the two-dimensional structure of the Rosette Nebula \citep{10.1093/mnras/staf1530}.  This work measured emission lines including H$\alpha$, [O\,{\sc iii}], [N\,{\sc ii}], and [S\,{\sc ii}] form the spectra with a resolution $R\sim4000$, thereby deriving the spatial distributions of line intensities as well as line ratios.  The findings reported in that work are largely consistent with those in this paper.  Specifically, the spatial distribution of line intensities reveals pronounced gradients that correspond to the optical abundance variations within the nebula.  Similarly, the observed gradients in line ratios and the inferred chemical abundances in that work exhibit the trends comparable to those documented in our study.

    \section{Data Accessibility}

        The fiber spectra, including those analyzed in this paper, from the LAMOST MRS-N survey \citep[see details about this project in][]{mrsn, mrsnHa, mrsnRV} have not yet been publicly released.

    \begin{acknowledgments}
        This work is supported by the National Natural Science Foundation of China (NSFC) GrantNos.12090041, 12090040, 2021YFA1600401,
        2021YFA1600400, 12273052, Nos.11733006, U1931109, 12003043 and the science
        research grants from the China Manned Space Project
        (No. CMS-CSST-2025-A14, CMS-CSST-2021-A04).

        We would like to thank Yunning Zhao for useful suggestions on data processing that helped to improve our work.

        Guoshoujing Telescope (the Large Sky Area Multi-Object Fiber Spectroscopic Telescope LAMOST) is a National Major Scientific Project built by the Chinese Academy of Sciences. Funding for the project has been provided by the National Development and Reform Commission. LAMOST is operated and managed by the National Astronomical Observatories, Chinese Academy of Sciences.
    \end{acknowledgments}

    \vspace{5mm}
    \facilities{LAMOST}

    \software{astropy,  
            tvwenger \citep{tvwenger}}

    \bibliographystyle{aasjournal}
    \bibliography{ms}

\end{document}